\documentclass[authoryear,12pt]{article}
\usepackage{amsmath,amsfonts,amssymb,epsfig,subfigure,color}
\usepackage{natbib}

\renewcommand{\le}{\leqslant}

\newcommand{\be}{\begin{equation}}
\newcommand{\en}{\end{equation}}

\renewcommand{\vec}[1]{\boldsymbol{#1}}

\begin{document}


\title{On the Rectilinear Shear of Compressible and Incompressible Elastic Slabs 
}


\author{M. Destrade$^a$, G. Saccomandi$^b$\\[12pt]
$^a$School of Mathematics, Statistics and Applied Mathematics, \\
  National University Of Ireland Galway, \\University Road, \\ Galway, Ireland\\[12pt]
 $^b$Dipartimento di Ingegneria Industriale, \\
 Universit\`{a} degli Studi di Perugia,
       \\ 06125 Perugia, Italy}
       
       \date{}
\maketitle


\begin{abstract}

 We review some pseudo-planar deformations for the equations of
incompressible isotropic nonlinear elasticity first introduced in 1985 by
Rajagopal and Wineman. We extend this class of deformations to
compressible isotropic and transverse isotropic materials, and also consider the influence of gravity.
We consider some new approximate solutions and we discuss the possible relevance of
such solutions to the understanding of the complex structure of the fields equations
of nonlinear elasticity,  using weakly nonlinear theories.

\end{abstract}
\bigskip


\newpage


\section{Introduction}


The starting point of our discussion is the 1985 paper by Rajagopal and Wineman entitled \textit{New Exact Solutions in Non-Linear Elasticity}. 
In that paper some new exact solutions to boundary value problems of nonlinear elasticity were established. 
These solutions constitute an important breakthrough point in the literature and to appreciate this fact, we have to review what was going around in Continuum Mechanics and in the field of nonlinear elasticity at the time. 

The origins of the modern theory of nonlinear elasticity must be related to the pioneering works of Rivlin, Green, and Adkins just after World War II. 
These fellows were the first to elaborate a reasonable and general notation to derive the right balance equations for nonlinear elasticity, and were the first to determine classes of exact solutions in nonlinear elasticity (for a summary of this earlier work see Green and Adkins, 1970)\footnote{To be more precise, Gabriella Armanni published a note in 1915, transmitted by Vito Volterra to the \emph{Nuovo Cimento} journal [\textbf{9} (1915) 427--447], with some radial solutions for a spherical nonlinear elastic solids, but this was an isolated episode.}. 

Then in 1955, Ericksen proved that homogeneous deformations are the only \textit{controllable} static deformations possible in every hyperelastic material. 
A controllable deformation is a deformation that is produced in a material by the application of surface tractions alone. 
A controllable deformation that can be effected in every homogeneous isotropic material is referred to as a \textit{universal solution}.  
Ericksen's result concerning universal deformations has had a profound influence on the development of nonlinear elasticity. 
For many years afterwards, there was ``the false impression that the only deformations possible in an elastic body are the universal deformations'' (Currie and Hayes, 1981). 
This is not exactly true and for a list of notable exceptions see for example the book by Ogden (1984). 
Currie and Hayes (1981) were right to point out that after some initial interest in the search for exact solutions in nonlinear elasticity, there followed a long period of inactivity concerning this enterprise. 
The paper by Currie and Hayes (1981) was the seed necessary to revitalize the search for possible solutions beyond universal solutions. The 1984 and 1985 papers by Rajagopal and Wineman were the first relevant new fruits.

In 1979, Ratip Berker proposed some new exact solutions for the Navier-Stokes equations. 
These are an infinite set of nontrivial solutions for an incompressible viscous fluid contained between the two parallel infinite plates rotating with constant angular velocity around the fixed normal axis. 
These flows are pseudo-plane flows of the first kind (Berker, 1963). 
Rajagopal was able to extend these solutions to a huge class of non-Newtonian fluids (Rajagopal, 1992) in an elegant note (Rajagopal, 1982). 
Rajagopal and Wineman (1984, 1985) considered the solid mechanics counterpart of the Berker solutions to obtain exact non-universal solutions for all incompressible isotropic elastic materials. 
This family of deformations is then generalized in (Rajagopal and Wineman, 1985) by considering that the bottom and top plates rotate with different angular velocities. 
These deformations are admissible solutions for all Mooney-Rivlin incompressible elastic materials.

Rajagopal has been a fine reader and connoisseur of the Berker encyclopedic article (Berker, 1963) where the pecularities of pseudo-plane flows are discussed in great detail.
Moreover, he established the formal analogy between the Navier-Stokes equations and the equations determining admissible deformations for neo-Hookean solids, and thus turned the Berker article into a formidable source of inspiration for new classes of solutions in nonlinear elasticity. 

To illustrate why pseudo-plane deformations constitute an interesting class of potential solutions in nonlinear elasticity, we consider some quantitative details. 
With material and spatial rectangular Cartesian coordinates $(X,Y,Z)$ and $(x,y,z)$, respectively, a \emph{pseudo-plane deformation} is a deformation of the form
\be \label{i1}
x=x(X,Y,Z), \quad y=y(X,Y,Z), \quad z=Z.
\en
An interesting way to generate such a deformation is to consider a plane deformation solution which depends on a certain number of arbitrary constants, $a, b, c, \ldots$, say, and to replace these constants with functions of $Z$. 
For example, by going from
\begin{equation}
 x=x(X, Y; a, b, c), \qquad 
 y=y(X, Y; a, b, c), \qquad
 z=Z,  
 \end{equation}
 to 
 \begin{equation}
 x=x(X, Y; a(Z), b(Z), c(Z)), \quad
 y=y(X, Y; a(Z), b(Z), c(Z)), \quad
 z=Z. 
\end{equation}
This idea was independently considered also by Hill and Shield in 1986 but only for neo-Hookean materials. 
It is in Rajagopal's work that it has resulted in many new exact solutions of non-linear elastic layers. 
The Berker solution is reconsidered also in (Fosdick et al, 1986); 
a non-uniform extension of a slab of Mooney-Rivlin material is considered in (Rajagopal et al., 1986); 
the rectilinear deformation of a general incompressible slab is considered in (McLeod et al., 1988); 
and not only deformations but also motions of elastic slabs are considered in (Carroll and Rajagopal, 1986). 
For pseudo-plane deformations and motions of the second kind (Berker, 1963), we refer to (Hayes and Rajagopal, 1992) and (Horgan and Saccomandi, 2003). 
In the paper (Rajagopal, 1984), the Berker (1979) solution is generalized to the case where the Newtonian flow is contained between the two parallel infinite porous plates. 
This suggests a generalization of \eqref{i1} such that
\be \label{i2}
x=x(X,Y,Z), \quad y=y(X,Y,Z), \quad z=\lambda(Z).
\en
This class of deformation is used in (Saccomandi, 2005) to obtain new exact solution for neo-Hookean solids. 

Pseudo-planar solutions of the Navier-Stokes equations have not been the only inspiring source of analogies to obtain new solutions in nonlinear elasticity from existing solutions in fluid mechanics. 
By considering the celebrated Jeffrey and Hamel convergent and divergent flows in intersecting  planes (Berker, 1963), Rajagopal and co-workers have produced interesting deformations admissible in nonlinear elasticity (Fu et al., 1990; Tao and Rajagopal, 1990; Rajagopal and Carroll, 1992; Rajagopal and Tao, 1992; McLeod and Rajagopal, 1999). 
This class of deformations has also been considered by Klingbiel and Shield (1962), but Rajagopal has been able to reverse the coin of Continuum Mechanics to transfer his fluid mechanics expertise to solid mechanics. 
 
The aim of this note is to push a little further the ideas of Rajagopal and investigate what happens to pseudo-planar solutions in the compressible case of nonlinear isotropic and anisotropic elasticity.
Hence in Section \ref{The-compressible-case}, we use \eqref{i2} to study rectilinear shear deformations coupled to an axial deformation of an elastic slab in the framework of compressible materials. 
We show that if we take into account the weight of the elastic slab, then the rectilinear shear solutions are non-trivial. 
It is usual in nonlinear elasticity theory to assume that the weight of an elastic structure is negligible when compared to elastic forces, but here we find that for certain fields of application such as geophysics, this is not necessarily the case, especially for thick layers.  
Then for incompressible materials (Section \ref{The-incompressible-case}), we derive a universal relation that went unnoticed by McLeod et al. (1988) and which proves very useful in simplifying the analysis of the problem (For universal relations in the framework of rectilinear shear see (Saccomandi, 1996)). 
We use this class of deformations further in Section \ref{anisotropic} to generalize some results to the case of the rectilinear shear of a transversely isotropic elastic material. 
In (Destrade et al., 2009) we considered orthotropic materials were the fibers are arranged in a special plane; here, the fibers may have a general orientation in space.  

In order to make progress and obtain exact solutions in nonlinear elasticity, a special constitutive assumption must be enforced. 
Indeed only Ericksen's universal solutions are valid irrespective of the choice of the strain energy density function. 
Sometimes the constitutive assumptions that have been considered in the literature were followed only for reasons of  mathematical convenience. 
This approach has generated a huge number of strain energy density functions that have no real connection with experimental evidence.  
Here we use a different approach, using the so called \textit{wealkly nonlinear theory of elasticity}. 
The strain energy associated with the classical linear theory of elasticity is of second order in the Green strain; here to investigate the non-linear behavior of elastic materials, we use strain energies that are of the third and fourth order in the strain. 
In such a way we are able to bring out some interesting general features of nonlinear elastic materials. 
For example, we show in Section \ref{anisotropic} that the coupling among the various mode of deformations occurs at lower order of non-linearity for anisotropic materials than for isotropic materials.   


\section{Governing equations}  

  
We call $\vec{x}(\vec{X})$ the current position of a particle which was located at $\vec{X}$ in the reference configuration. 
Two kinematic quantities associated with this motion are the deformation gradient and the left Cauchy-Green strain tensor,
\be
\vec{F} = \partial \vec{x} / \partial \vec{X}, \qquad \vec{B} = \vec{FF}^{T},
\label{01}
\en
respectively.
In Section \ref{isotropic}, we consider hyperelastic, isotropic materials, and so we introduce the strain energy density  $W = W(I_{1},I_{2}, I_{3})$, where $I_{1}$ and $I_{2}$ are the first and second principal invariants of $\vec{B}$, respectively, given by
\be
I_{1}=\text{tr} \ \vec{B}, \qquad I_{2}= \textstyle{\frac{1}{2}} \left[I_1^2-\text{tr}(\vec{B}^2) \right],
\label{02}
\en
and $I_{3}=J^2$ where $J=\det \vec{F}$.
The general representation formula for the Cauchy stress tensor $\vec{T}$ reads
\begin{equation}
\vec{T}=\beta_0 \vec{I}+\beta_1 \vec{B} + \beta_{-1} \vec{B}^{-1},
\label{1}
\end{equation}
where 
\be
\beta_0 = 2J^{-1} \left[I_2W_2+I_3W_3 \right], \qquad 
\beta_1= 2J^{-1}W_1, \qquad 
\beta_{-1}=-2J W_2,
\en
and $W_1 \equiv \partial W / \partial I_1$, $W_2 \equiv \partial W/\partial I_2$, $W_3 \equiv \partial W/\partial I_3$.

If the material is incompressible then the only admissible deformations are isochoric, i.e. $J=1$ at all times, so that $I_{3} \equiv 1$, and $W=W(I_1, I_2)$ only. 
In this case,
\begin{equation}
\vec{T}=-p\vec{I}+2 W_1 \vec{B} - 2 W_2 \vec{B}^{-1},
\label{2}
\end{equation}
where $p$ is the yet indeterminate Lagrange multiplier introduced by the constraint of incompressibility. 

The link with linear elasticity is made by defining $\mu$, the infinitesimal shear modulus, as
\be \label{mu}
\mu = 2(W_1 + W_2)|_{I_1=I_2=3, I_3=1}.
\en

In Section \ref{anisotropic}, we consider a special class of hyperelastic \emph{anisotropic} materials, namely transversely anisotropic  materials, i.e. with a single preferred direction.
We call $\vec{M}$ a unit vector along this direction in the reference configuration and we introduce the anisotropic invariant
\be
I_4 \equiv \vec{M} \cdot \vec{C M}-1 = \vec{m} \cdot \vec{m}-1,
\en
where $\vec{C}=\vec{F}^{T}\vec{F}$ is the right Cauchy-Green strain tensor, and  $\vec{m}=\vec{F}\vec{M}$.

We restrict our attention to compressible, transversely isotropic materials for which  $W=W(I_1, I_2, I_3, I_4)$. 
Then the constitutive equation \eqref{1} is replaced by
\begin{equation}
\vec{T}=\beta_0 \vec{I}+\beta_1 \vec{B} + \beta_{-1} \vec{B}^{-1}+\beta_{4}\vec{m} \otimes \vec{m},
\label{1bis}
\end{equation}
where $\beta_4=2J^{-1}W_4$, $W_4=\partial W/\partial I_4$.
For incompressible, transversely isotropic materials with $W=W(I_1, I_2, I_4)$, the representation formula \eqref{2} is replaced with
\begin{equation}
\vec{T}=-p\vec{I}+2 W_1 \vec{B} - 2 W_2 \vec{B}^{-1}+2W_4 \vec{m} \otimes \vec{m}.
\label{2bis} 
\end{equation}

 
The balance equation of linear momentum, \emph{in the absence of body forces}, reads
\begin{equation}
\text{div}\, \vec{T} =\vec{0}.  \label{3}
\end{equation} 
If we consider the presence of body forces such as gravity\footnote{In nonlinear elasticity we do not know of exact solutions taking into account the presence of gravity. 
In linear elasticity and in incremental elasticity, a classical exact solution taking gravity into account is the stretching of a bar by its own weight,  see the book by Timoshenko and Goodier (1982) and the article by Fosdick and Shield (1963), respectively. 
}, for example, then we have
\begin{equation}
\text{div}\, \vec{T} +\rho \vec{g}=\vec{0},  \label{4}
\end{equation} 
where $\rho \vec{g}$ is the weight per unit of volume of the body in the current configuration and $\rho$ is the current mass density. 
By the conservation of mass, 
\be
\rho_0=J\rho,
\en
where $\rho_0$ is the mass density in the reference configuration.





\section{Rectilinear shear and axial stretch}
\label{isotropic}


Among the class of deformations in \eqref{i2} a special status is detained by the deformations
\begin{equation}  
x=A X+ B Y + f(Z),\qquad y=C X + D Y+g(Z), \qquad z=h(Z),  \label{p1}
\end{equation}
where $A$, $B$, $C$, $D$ are constants and $f$, $g$, $h$ are functions of $Z$ alone.
These deformations consist of two shearing deformations $f(Z)$ and $g(Z)$ in the $Z$-direction, combined to an inhomogeneous stretch $h(Z)$ along the $Z$-axis, and superimposed on a plane homogeneous deformation.

The class of deformations \eqref{p1} is not universal, but it reduces the balance equations to a system of three ordinary differential equations for any choice of the strain energy density.
Moreover, at $A=D=1$, $B=C=0$, it is the static counterpart to the usual longitudinal and transverse wave motions propagating along the $Z$-axis and polarized along the three reference axes. 
For the choices $h(Z)=\lambda Z$, $A=D=\lambda^{-1/2}$, $B=C=0$, where $\lambda$ is a constant, we obtain an isochoric deformation first considered in a dynamical context by Carroll (1967), and then by Rajagopal and co-workers. 
In those papers, the material was incompressible, while here we are considering both compressible and incompressible materials. 
Another interesting case occurs for the choices
\be \label{p111}
A=D=\cos \Omega, \qquad B=-\sin \Omega, \qquad C=\sin \Omega, \qquad h(Z)=\lambda Z,
\en
where $\Omega$ and $\lambda$ are constants.
Here we recover the field studied by Rajagopal and Wineman (1984, 1985), describing the deformation of a nonlinear elastic layer contained between two infinite parallel rigid plates, each of which undergoes the same finite rotation.

Considering now the class of deformations \eqref{p1} in all generality, we find
\be
[\vec{F}]_{ij} = 
\begin{bmatrix}
A & B &  f'
\\
C & D & g' 
 \\
 0 & 0 & h'
 \end{bmatrix}, \qquad
 [\vec{F}^{-1}]_{ij} =  J^{-1}
\begin{bmatrix}
 Dh'& -Bh' &  Bg'-Df'
\\
-Ch' & Ah' &  Cf'-Ag' 
 \\
0  & 0 & AD-BC
 \end{bmatrix}.
\en
where $J = (AD-BC)h'$ and the prime denotes differentiation with respect to $Z$.
Also, 
\be
[\vec{B}]_{ij} = 
\begin{bmatrix}
A^2 + B^2 + f'^2 & AC + BD + f' g' &  f'h'
\\
AC + BD + f' g' & C^2 + D^2 + g'^2 & g' h'
 \\
 f'h' & g' h' & h'^2
 \end{bmatrix}, 
\en
and
\begin{align}
 &[\vec{B}^{-1}]_{11} =  J^{-2} (C^2+D^2)h'^2, \notag \\
 &[\vec{B}^{-1}]_{12} =  -J^{-2}(AC+BD)h'^2 = [\vec{B}^{-1}]_{21}, \notag \\
 &[\vec{B}^{-1}]_{13} =  J^{-2}[(AC+BD)g'-(C^2+D^2)f']h' = [\vec{B}^{-1}]_{31}, \notag\\
&[\vec{B}^{-1}]_{22} =  J^{-2} (A^2+B^2)h'^2, \notag \\ 
&[\vec{B}^{-1}]_{23} =  J^{-2} [(AC+BD)f'-(A^2+B^2)g']h' = [\vec{B}^{-1}]_{32}, \notag \\ 
&[\vec{B}^{-1}]_{33} =  J^{-2}[(Bg' - Df')^2 + (Ag' - Cf')^2 + (AD - BC)^2].
\end{align}
The principal invariants follow as
\begin{align}
& I_1 = A^2 + B^2 + C^2 + D^2 + f'^2 + g'^2 + h'^2, \notag \\
& I_2 = (AD - BC)^2 + (Df'-Bg')^2 + (Ag' - Cf')^2 + (A^2 + B^2 + C^2 + D^2)h'^2, \notag \\
& I_3 = (AD - BC)^2 h'^2.
\end{align}
 
Now we may compute the Cauchy stress components for compressible solids from \eqref{1} as
\begin{align}
&T_{11}=\beta_0+\beta_1(A^2+B^2+f'^2) + \beta_{-1}J^{-2}(C^2+D^2)h'^2,\notag
\\
&T_{22}=\beta_0+\beta_1(C^2+D^2+g'^2)+\beta_{-1}J^{-2}(A^2+B^2)h'^2,\notag
\\
&T_{33}=\beta_{0}+\beta_1h'^2+\beta_{-1}J^{-2}[(Bg'-Df')^2+(Ag'-Cf')^2+(AD-BC)^2],
\notag \\
& T_{12}=\beta_1 (AC + BD + f'g') - \beta_{-1}J^{-2}[(AC+BD)h'^2, \notag 
\notag \\
& T_{13}=\beta_1 f'h'+\beta_{-1}J^{-2}[(AC+BD)g'-(C^2+D^2)f']h', \notag 
\\
& T_{23}=\beta_1 g'h'+\beta_{-1}J^{-2} [(AC+BD)f' - (A^2+B^2)g']h'.  
\end{align} 
For incompressible solids, we may use the same formulas, by replacing $\beta_0$ with $-p$, $\beta_1$ with $W_1$, and $\beta_{-1}$ with $W_2$.


\subsection{The compressible case}
\label{The-compressible-case}


Now we write down the balance equations \emph{in the absence of body forces}, equations \eqref{3}.
They reduce to 
\be \label{p2}
T_{13}'=0, \qquad T_{23}' = 0, \qquad T_{33}' = 0, 
\en
and they may readily by integrated as 
\be \label{homog}
T_{13}=\tilde{k}_1, \qquad T_{23}=\tilde{k}_2, \qquad T_{33}=\tilde{k}_3,
\en
where $\tilde{k}_1$, $\tilde{k}_2$, $\tilde{k}_3$ are constants. 
This is a non-homogeneous quadratic system of three equations with constant coefficients for the three unknowns $f'$, $g'$, $h'$. 
Once solved, it yields $f'$, $g'$, $h'$ as constants, showing that the only possible solutions are the trivial homogeneous solutions. 
This is in contrast with the dynamic counterpart of these deformations (see Destrade and Saccomandi, 2006). 

Next, consider the case where \emph{the gravity body force is present}, so that $\rho \vec{g}$ has component $- \rho g = - \rho_0 J^{-1} g$ along the $Z$-direction.  
Then the balance equations \eqref{4} reduce to 
\be \label{p3}
T_{13}'=0, \qquad T_{23}' = 0, \qquad T_{33}' = \rho_0 (AD - BC)^{-1}g, 
\en
and they may readily by integrated as 
\be \label{compr}
T_{13}=\tilde{k}_1, \qquad T_{23}=\tilde{k}_2, \qquad T_{33}= \rho_0 (AD - BC)^{-1}gZ + \tilde{k}_3,
\en
where $\tilde{k}_1$, $\tilde{k}_2$, $\tilde{k}_3$ are constants. 
In contrast to \eqref{homog}, this differential system admits \emph{non-homogeneous} solutions. 

Take the case $A=D=1$, $B=C=0$.
Then, the governing equations \eqref{compr} reduce greatly, to 
\begin{align} \label{p4} 
&[\beta_1-\beta_{-1}(h')^{-2}]f'h'=\tilde{k}_1,  \notag
\\ 
&[\beta_1-\beta_{-1}(h')^{-2}]g'h'=\tilde{k}_2,  \notag
\\ 
&\beta_{0} + \beta_1 (h')^2 + \beta_{-1} (h')^{-2}[1 + (f')^2 + (g')^2] = \rho_0 g Z + \tilde{k}_3. 
\end{align}

To set down a boundary value problem, we consider a slab of finite thickness $H$ along $Z$ and of infinite extent otherwise. 
Then we scale the lengths with respect to $H$ and the $\beta$'s with respect to $\mu$, the infinitesimal shear modulus  defined in \eqref{mu}. 
The non-dimensional version of the system \eqref{p4} is thus written in the domain $0 \le Z \le1$, as
\be \label{p5}
Q_1 f' = k_1, \qquad Q_1 g' = k_2, \qquad (Q_1 + Q_2) h' = (\rho_0 g H/\mu) Z + k_3,
\en
where
\be
Q_1 = 2(W_1 + W_2)/\mu, \quad 
Q_2 = 2(W_2 + W_3)/\mu, \quad 
k_i = \tilde{k}_i/\mu.
\en 
Notice that when $f' \neq 0$, $g' \neq 0$, we can deduce from \eqref{p5}$_{1,2}$ that $f'$ is proportional to $g'$, thus reducing the dimension of the system of ordinary differential equations.
Notice also that we can differentiate \eqref{p5} with respect to $Z$, to get the equivalent form of the governing equations,
\be \label{eqn}
\left(Q_1 f'\right)' = 0, \qquad \left(Q_1 g'\right)' = 0, \qquad \left[(Q_1 + Q_2) h'\right]' = (\rho_0 g H/\mu).
\en

It is possible to impose Dirichlet boundary conditions by prescribing the displacements at the bottom and the top of the slab, or Neumann boundary conditions by imposing the values of shear stresses $T_{12}$, $T_{13}$ and of the normal stress $T_{33}$ on the faces of the slab. 
Mixed boundary conditions are also possible.
 
Now we specialize the analysis to the Murnaghan strain-energy density:
\be \label{murn}
W = \dfrac{\lambda + 2 \mu}{8} J_1^2 + \dfrac{\mu}{2} J_2 + \dfrac{l+2m}{24}J_1^3 + \dfrac{m}{4} J_1J_2 + \dfrac{n}{8} J_3, 
\en
where $\lambda$ and $\mu$ are the Lam\'e coefficients of second-order elasticity, and $l$, $m$, $n$ are the Murnaghan third-order constants.
Here, $J_1$, $J_2$, $J_3$ are another set of independent invariants, related to the principal invariants of strain through:
\be
J_1 = I_1-3, \qquad
J_2 = 2 I_1 - I_2 - 3, \qquad
J_3 = I_3 - I_2 + I_1 - 1.
\en
For this strain energy,
\begin{align}
& Q_1 = 1 + \dfrac{\lambda+2\mu+m}{2\mu}J_1 + \dfrac{l+2m}{4\mu} J_1^2 + \dfrac{m}{2\mu}J_2, 
\notag \\
& Q_1 + Q_2 = \dfrac{\lambda+2\mu}{2\mu}J_1 + \dfrac{l+2m}{4\mu} J_1^2 + \dfrac{m}{2\mu}J_2.
\end{align}

To work out an explicit example, we take the deformation to be in the form $f(Z)= u(Z)$, $g(Z) = 0$, $h(Z) = Z + w(Z)$, for some functions $u$, $w$ of $Z$, so that the mechanical displacement is $\vec{x} - \vec{X} = [u, 0, w]^t$.
Then we find that 
\be
J_1 = 2 w' + (u')^2 + (w')^2, \qquad
J_2 = (u')^2, \qquad
J_3 = 0.
\en 
%
We also impose the displacements boundary conditions $f(0)=0$, $h(0)=1$, $h(1)=\ell$, equivalent to  
\be\label{bc1}
u(0)=0, \qquad w(0)=0, \qquad w(1) = \ell -1. 
\en


First of all we make the connection with the solution of \emph{linear elasticity}. 
There, $u=u_0$, $w=w_0$, say, where $u_0$ and $w_0$ are infinitesimal quantities, in the sense that $u_0^2$, $w_0^2$ and higher powers are negligible when compared to $|u_0|$ and $|w_0|$. 
At that order of approximation, \eqref{eqn} reduces to 
\be
u_0''=0, \qquad w_0''= \dfrac{\rho g H}{\lambda + 2 \mu},
\en
with solution:
\be \label{u0w0}
u_0=k_1 Z, \qquad w_0= \dfrac{\rho g H}{2(\lambda + 2 \mu)}(Z-1)Z + (\ell-1)Z,
\en
where the integration constant $k_1$ is to be determined later on, from the 
the stress boundary condition imposed on $T_{13}(1)$. 
It is clear that there is a limitation to this solution in the sense that the expression for $w_0$ must stay within the so-called ``elastic limit'', and this is not guaranteed for slabs with large thickness $H$. 
In real world applications where gravity plays a significant role (such as geophysics), it could well be the case that a non-linear correction is required to account for the effects of a large layer thickness.  

We now move on to the next order of elasticity theory, that is the one encompassed by the strain energy \eqref{murn}.
To obtain a solution for the boundary value problem, we use a simple perturbation method, by expanding the displacement as $u=u_0 +  u_1 + \ldots$, $w = w_0 + w_1 + \ldots$, say, in the same spirit as the method of successive approximations  introduced by Signorini (1949), and later studied by many authors (see for example, Lindsay (1985)). 
Here, $u_0$, $w_0$ are given by \eqref{u0w0}, $|u_1|$ and $|w_1|$ are of the same order as $u_0^2$ and $w_0^2$, and higher orders are neglected.
From \eqref{eqn} we obtain the following system of equations,
\begin{align}
& \left[u_1'+ \dfrac{\lambda+2\mu+m}{\mu} u_0'w_0'\right]'=0, \notag \\ 
& \left[w_1'+ \dfrac{\lambda+2\mu+m}{\mu}(u_0')^2 +  \dfrac{3 \lambda + 6\mu + 2l + 4m}{2\mu}(w_0')^2\right]'=0.
\end{align}
Clearly now, each component of the next order solution involves a combination of the longitudinal deformation (through $w_0$) and the shear deformation (through $u_0$).
Owing to the form \eqref{u0w0} of the lower order solutions, these equations reduce to
\begin{align} \label{u1w1}
& u_1'' +  \dfrac{\lambda + 2\mu + m}{\mu(\lambda+2\mu)} k_1 \rho g H=0, \notag \\[8pt] 
& w_1'' +  \dfrac{3 \lambda + 6\mu + 2l + 4m}{2\mu(\lambda+2\mu)} \rho g H
\left[ \dfrac{\rho g H}{2(\lambda + 2 \mu)}(2Z-1) + (\ell-1)\right] = 0.
\end{align}
Hence, the rectilinear shear is no longer a simple shear deformation. 
Here we have, according to \eqref{u1w1}$_1$, a quadratic variation of $u$ with $Z$, 
\begin{equation}
u= k_1 Z - \dfrac{\lambda+2\mu+m}{\lambda + 2 \mu} \dfrac{\rho g H}{2\mu}k_1 Z^2,
\end{equation}
and, according to \eqref{u1w1}$_2$, a cubic variation of $w$ with $Z$, which we do not reproduce for brevity. 

In his treatment of gravity seismic waves Biot (1940) gives $\rho g /\mu$ as being of the order of $2\times 10^{-6}$ m$^{-1}$ for the Earth. 
This means that the non-linear correction above is of the same order as the linear solution when $H \sim 1000$ km, or about one-sixth of the Earth radius. 
The value chosen by Biot is typical of  common rocks such as granite, but it is unlikely to be representative of material properties at such depths. 
For softer grounds such as mud or sediments, $\rho g /\mu$ is much larger, of the order of $2.5 \times 10^{-4}$ m$^{-1}$ say (Holzer et al., 2005), indicating that $H \sim 8$ km for the non-linear correction to be of the same order as the linear solution. 
At such depth however, consolidation has occurred and the medium is much stiffer than near the surface.
The conclusion of these estimates is that on physical grounds, we do not need to push the Murnaghan expansion to fourth order, because third-order non-linear effects are simply a very small correction to the linear solution.

\color{black}
Nonetheless, simple shear deformations are an important component of most \emph{geophysical applications}. 
For example, they form the basis of an explanation of the folding phenomenon \citep{Manz79}.
Similarly, the determination of shear strength, viscosity, and internal friction data for deep crust and upper mantle rock and mineral analogues under geophysically realistic conditions of very high temperatures and pressures is required in order to interpret earthquake origins, seismic signal generation, and explosions (see \cite{Okumura10} and references therein) as well as the rheological evolution of the microstructure of mantle materials \citep{Karato98}. 
It is clear that in order to have a realistic geophysical picture, it is necessary to conduct a numerical investigation of equations \eqref{eqn}, and to include the effects of very high temperatures and pressures. 
We note that under these circumstances, the constitutive properties of geophysical materials should quite softer than those indicated by \cite{Biot40}, leading to a critical $H$ much more smaller than the one we obtained above in our crude estimation.  
This means that the effect of gravity, and the corresponding inhomogeneous correction predicted by the Murnaghan theory, might both have to be taken into account after all.
  
\color{black}


\subsection{The incompressible case}
\label{The-incompressible-case}


Here we go back to the incompressible case, a case already examined by McLeod, Rajagopal, and Wineman (1988). 
The aim of this subsection is to show that the use of a universal relation reduces the problem to a simple formulation.

In the incompressible  case the deformation \eqref{p1} must be isochoric.
Here, we follow the choice of Carroll (1967) and of Mc Leod et al. (1988), by taking $h(Z) = \lambda Z$, $A=D=\lambda^{-1/2}$, $B=C=0$.
We study the equations of equilibrium \emph{in the absence of body forces}.
In contrast with the compressible case, non-trivial (non-homogeneous) solutions are possible, due to the effects of the Lagrange multiplier $p$. 
Indeed, the first two equations of equilibrium lead to $\partial p /\partial x = c_1$, $\partial p /\partial y = c_2$, where $c_1$ and $c_2$ are constants. 
Assuming that $c_1 c_2 \ne 0$ excludes the possibility of trivial homogeneous deformations. 
Then, the determining equations for the rectilinear shear deformation $f(Z)$ and $g(Z)$ are   
\be \label{p02}
Q_1 f'=c_1 Z+ {k}_1, \qquad Q_1 g'=c_2 Z+ {k}_2, 
\en
while the third equation, corresponding to \eqref{p3},  determines the unknown Lagrange multiplier $p$.
We recall that gravity  is a conservative external force: it would change the form of $p$  but would not influence the shearing deformations, which is why we did not include it in this section.


We note that McLeod et al. (1988) did not make use of the following \emph{universal relation}
\be
g'=\frac{c_2 Z+ {k}_2}{c_1 Z+ {k}_1} f'. 
\en
obtained directly from \eqref{p02}.
With this universal relation, the discussion of boundary value problems is greatly simplified, as we now see for the most general case of fourth-order incompressible elasticity,
\be \label{incomp4}
W = \mu \; \text{tr}\left(\vec{E}^2\right)  + \frac{\mathcal{A}}{3} \;\text{tr}\left(\vec{E}^3\right)   + \mathcal{D} \; \left(\text{tr} (\vec{E}^2)\right)^2,
\en
where $\vec{E} = (\vec{F}^T\vec{F} - \vec{I})/2$ is the Green strain tensor, and $\mu$, $\mathcal{A}$, and $\mathcal{D}$ are second, third-, and fourth-order elasticity constants, respectively (Hamilton et al., 2004; Ogden, 1974).
For convenience, we work with the equivalent strain energy density 
\be
W(I_1, I_2) = C_{10}(I_1-3) + C_{01}(I_2-3) + C_{20}(I_1-3)^2,
\en
where $C_{10}$, $C_{01}$,  and $C_{20}$ are constants. 
At the same degree of approximation in the Green strain $\vec{E}$, it covers \eqref{incomp4} with the identifications (Destrade et al., 2010):
\be \label{connections}
\mu = 2(C_{10} + C_{01}), \qquad 
\mathcal{A} = -8(C_{10} + 2C_{01}), \qquad
\mathcal{D} = 2(C_{10} + 3C_{01} + 2C_{20}).
\en
and at $C_{20}=0$, it also covers the Mooney-Rivlin case of McLeod et al. (1988).
Then the determining equations \eqref{p02} reduce to a coupled system for the shear modes:
\begin{align}
& [\alpha + \beta (f'^2+g'^2)] f'= c_1 Z+ {k}_1, \notag \\ 
& [\alpha + \beta(f'^2+g'^2)] g'= c_2 Z+ {k}_2,
\end{align}
where the scalar $\alpha$ and $\beta$ are defined as
\be
\alpha = 1 + (\lambda^2 + 2\lambda^{-1} - 3)\beta,
\qquad
\beta = 2\dfrac{C_{20}}{C_{10} + C_{01}}.
\en

We consider the following boundary conditions $f(0)= f(1)=0$ and $g(0)= g(1)=0$.
Using symmetry arguments (see Saccomandi (2004) for details), it may then be shown that ${k}_1$ and ${k}_2$ are such that
\begin{align}
& [\alpha + \beta (f'^2+g'^2)] f' = c_1 \left(Z - \textstyle{\frac{1}{2}} \right), 
\notag \\
& [\alpha + \beta (f'^2+g'^2)] g'=c_2 \left(Z - \textstyle{\frac{1}{2}} \right).
\end{align}
Using now the universal relation, which reads $f'= (c_1/c_2) g'$, the problem is reduced to that of a classical rectilinear shear deformation
\be \label{zorro}
\alpha (g') + \beta\left(1+ \dfrac{c_1^2}{c_2^2}\right)(g')^3 =c_2 \left(Z - \textstyle{\frac{1}{2}} \right), \qquad g(0)=g(1)=0.
\en
Note that at $C_{20}=0$ we recover the solution of McLeod et al. (1988) for the Mooney-Rivlin material: 
\be
g(Z) = \dfrac{c_2}{2}Z(Z-1).
\en

Now we perform a perturbation scheme, as in the previous section. 
Hence we take $g$ in the form $g = v_0 + v_1 + v_2 + \ldots$, say, where $v_0$ is infinitesimal, $v_1$ is of order $v_0^2$, $v_2$ is of order $v_0^3$, etc., and we take the stretch in the form $\lambda = 1 + e + e^2 + \ldots$, where the elongation $e$ is infinitesimal.
Then we find that the quantity $\alpha$ expands as $\alpha = 1 + 3\beta e^2 + \ldots$ and $\beta$ remains the same.
We also find that the single ordinary differential equation \eqref{zorro} gives in turn,
\be
v_0'= c_2 \left(Z - \textstyle{\frac{1}{2}} \right), \qquad \text{so that} \quad 
v_0(Z) = \dfrac{c_2}{2}Z(Z-1),
\en
then 
\be
v_1'= 0, \qquad \text{so that} \quad 
v_1= 0,
\en
showing that we were right to push the expansion of $W$ to the fourth-order,
and finally, the non-linear correction,
\be
v_2' = - \beta\left[ \left(1+\dfrac{c_1^2}{c_2^2}\right) (v_0')^3 + 3e^2 (v_0')\right].
\en


We thus find the solution for $g$ as
\be 
g(Z)=\dfrac{c_2}{2}Z(Z-1)\left\{ 1 -\dfrac{\beta}{4}\left[(c_1^2 + c_2^2)(2Z^2 - 2Z + 1) + 12e^2 \right] + \ldots \right\},
\en 
and for $f$, by integrating the universal relation, as $f(Z) = (c_1/c_2)g(Z)$.
Thanks to the successive approximation solutions, and to the universal relation, we have been able to understand the role of higher order non-linearity and the coupling between the two shear components of the pseudo-planar deformation.


\section{Rectilinear shear in anisotropic materials.}
\label{anisotropic}


Let us consider a material reinforced with fibers aligned with the direction $\vec{M}=M_1 \vec{E}_1+ M_2 \vec{E}_2 +M_3 \vec{E}_3$, where the constants $M_1$, $M_2$, $M_3$ are of the form
\be \label{an0}
M_1=\sin \theta \sin \phi, \quad M_2=\sin \theta \cos \phi, \quad M_3=\cos \theta,
\en
with $\theta$, the elevation and $\phi$, the azimuthal angle. 

We restrict our attention to the incompressible case and the isochoric motions
\be
x = X + f(Z), \qquad y = Y + g(Z), \qquad z=Z.
\en
Then $\vec{m} = \vec{FM}$ has components
\be
[\vec{m}]_i=\left[M_1+f' M_3,  M_2 + g' M_3, M_3 \right]^T,
\en
so that 
\be
I_4 = \vec{m \cdot m}-1 = \left[(f')^2 + (g')^2\right] M_3^2 + 2(f' M_1+ g' M_2)M_3.
\en
 
The representation formula \eqref{2bis} gives the following shear stress components
\begin{align} \label{an1}
& T_{13}=2(W_1+ W_2)f'+ 2W_4 M_3 (M_1+f'M_3),
\notag \\
& T_{23}=2(W_1+ W_2)g'+ 2W_4 M_3 (M_2+g'M_3).
\end{align}
In passing we note that although $W$ may be linear in $I_1$ and $I_2$ (as in the Mooney-Rivlin model), it may not be linear in $I_4$, because otherwise there would be non-zero shear stresses in the reference configuration.
The balance equations \eqref{3} reduce now to 
\be
-p_{x} + T_{13}'(Z) = 0, \qquad  
-p_{y} + T_{23}'(Z) = 0, \qquad 
T_{33}'(Z) = 0.
\en
The third equation here can be solved by an appropriate choice for $p$. 
The first and second equations are compatible when
\be \label{an3}
T_{13}' = 2c_1, \qquad T_{23}' = 2c_2,   
\en
where $c_1$ and $c_2$ are constants.
Integrating, we obtain
\begin{align} \label{an4}
& (W_1+ W_2)f'+ W_4 M_3 (M_1+f'M_3) = c_1 Z + c_3, 
\notag \\
& (W_1+ W_2)g'+ W_4 M_3 (M_2+g'M_3) = c_2 Z + c_4,
\end{align}
where $c_3$, $c_4$ are constants. 

First, consider the case of a single shearing deformation: $f' \ne 0$, $g'=0$, say.
Then the second equation above results in 
\be
W_4 M_3 M_2=c_2 Z+c_4,
\en
which cannot be satisfied in general because $W_4$ is not constant. 
Therefore a shearing deformation in a single direction is not compatible with a general  orientation of the fiber arrangement. 
Next, consider the combination of two shearing deformations $f' g' \ne 0$.
Clearly then, the governing equations \eqref{an4} are compatible with a 
general fiber distribution in space.
This observation also holds when the angles $\theta$ and $\phi$ in \eqref{an0} depend on $Z$: $\theta=\theta(Z)$ and $\phi=\phi(Z)$.  
This has of course important biomechanical implications, because it is well established that collagen fiber bundles change their orientation in soft tissues (Holzapfel and Ogden, 2009).
 
 Now consider the standard reinforcing model for the Mooney-Rivlin solid,
\be
W = C_{10}(I_1 - 3) + C_{01}(I_2 - 3) + \textstyle{\frac{1}{4}} \gamma (I_4-1)^2,
\en
where $C_{10}$, $C_{01}$, $\gamma$ are constants.
Then the governing equations \eqref{an4} read
\begin{align} \label{an4bis}
& \mu f'+ \gamma[(f'^2+g'^2)M_3^2+(M_1f'+M_2g')M_3] M_3 (M_1+f'M_3)=c_1 Z+c_3, 
\notag \\
& \mu g'+ \gamma[(f'^2+g'^2)M_3^2+(M_1f'+M_2g')M_3] M_3 (M_2+g'M_3)=c_2 Z+c_4,
\end{align}
where $\mu = 2(C_{10}+C_{01})$.
The isotropic version (at $M_3=0$) of this system is a non-homogeneous \emph{linear} decoupled system of equations for $f'$ and $g'$, which may be solved for one shear deformation independently of the other. 
Clearly, the anisotropy gives a  coupled, non-linear (cubic) system of equations for the shear deformations.  
We may say that  anisotropic materials are more sensitive than isotropic materials to nonlinear effects.
\color{black}
For a detailed investigation of this type of sensitivity, we refer to the works of \cite{Merodio07} and \cite{Destrade09}.
\color{black}



 \color{black}
 

\section{Concluding remarks}

We revisited a class of deformations previously studied into details by Rajagopal and coworkers over several papers.
This class of deformation is the pseudo-planar version of some homogeneous deformations, composed of two inhomogeneous rectilinear shears superimposed 
onto an homogenous deformation. 

First, we extended the class of deformations from incompressible to compressible materials, by introducing the possibility an inhomogeneous axial stretch. 
In that case, the class of deformations is clearly not isochoric. In the presence of a body force such as gravity the solutions to the problem under investigation are not trivial i.e. we have non-homogeneous deformations. 
In the linearized case, the axial stretch and the rectilinear shear functions are uncoupled, but 
in the weakly nonlinear theories of elasticity, the situation becomes more intricate.  
At the order above linear elasticity (i.e.  third-order elasticity), we saw that the rectilinear shears are coupled to the axial stretch, although the two rectilinear shears do not interact with each other. 
Presumably, we would find a fully coupled situation at the next order (fourth-order elasticity), but it was not necessary to go that far, because we computed that those effects are negligible for commonly used values of physical constants.
This situation nonetheless revealed an important feature of the structure of the equations of nonlinear compressible elasticity.
Indeed, volume variations are very important in the general nonlinear theory.  Therefore a special mathematical compressible model which supports isochoric deformations must be seen as the exception rather than the rule, and must be handled with care because it might give false informations about the physics of compressible elastic materials.

In the incompressible case we noticed an important universal relation, which allowed us to simplify the qualitative analysis of the determining equations for the rectilinear shear unknowns. We provide a simple approximate solution of this problem.

Then we considered what happens in the framework of anisotropic materials, where lower-order coupling appears between the shearing deformations. 
This coupling is due to the presence of preferred fiber directions and may thus have interesting underpinnings in biomechanical applications.
Therefore it is possible to learn a lot from a simple semi-inverse problem about the complex structure of nonlinear elasticity.

We conclude with an open problem. 
For the incompressible Mooney-Rivlin material, Rajagopal and Wineman (1985) found solutions to the balance equations within the class described by \eqref{p1}, with \eqref{p111} in force when $\Omega=\psi Z+\psi_o$, i.e. when the angles of rotation of the two bounding planes of the slab are different. 
We ask whether it is possible to find a special class of \emph{compressible} materials for which this is possible?


\section*{Acknowledgements}


This work is supported by a Senior Marie Curie Fellowship awarded by the Seventh Framework Programme of the European Commission to the first author.




\end{document}